%% file: main.tex
\documentclass[sigconf]{acmart} 
\usepackage{longtable}
\usepackage{tabularray}
\usepackage{pdfpages}
\usepackage{enumitem}
\usepackage{soul}
\AtBeginDocument{%
  \providecommand\BibTeX{{%
    \normalfont B\kern-0.5em{\scshape i\kern-0.25em b}\kern-0.8em\TeX}}}

\setcopyright{acmlicensed}
\copyrightyear{2018}
\acmYear{2018}
\acmDOI{XXXXXXX.XXXXXXX}

\acmISBN{978-1-4503-XXXX-X/18/06}

\copyrightyear{2026}
\acmYear{2026}
\setcopyright{cc}
\setcctype{by}
\acmConference[CHI '26]{Proceedings of the 2026 CHI Conference on Human Factors in Computing Systems}{April 13--17, 2026}{Barcelona, Spain}
\acmBooktitle{Proceedings of the 2026 CHI Conference on Human Factors in Computing Systems (CHI '26), April 13--17, 2026, Barcelona, Spain}
\acmPrice{}
\acmDOI{10.1145/3772318.3790690}
\acmISBN{979-8-4007-2278-3/2026/04}

\begin{document}

\title{From Clicks to Consensus: Collective Consent Assemblies for Data Governance}
\author{Lin Kyi}
\affiliation{%
   \institution{Max Planck Institute for Security and Privacy}
  \city{Bochum}
  \country{Germany}
 }

\author{Paul Gölz}
\affiliation{%
   \institution{Cornell University}
  \city{Ithaca}
  \country{USA}
 }

 \author{Robin Berjon}
\affiliation{%
   \institution{Independent, Supramundane Agency}
  \city{Brussels}
  \country{Belgium}
 }

 \author{Asia J. Biega}
 \affiliation{%
   \institution{Max Planck Institute for Security and Privacy}
   \city{Bochum}
   \country{Germany}
 }

\renewcommand{\shortauthors}{Kyi et al.}

\begin{abstract}
Obtaining meaningful and informed consent from users is essential for ensuring autonomy and control over one's data. Notice and consent, the standard for collecting consent, has been criticized. While other individualized solutions have been proposed, this paper argues that a collective approach to consent is worth exploring. First, individual consent is not always feasible to collect for all data collection scenarios. Second, harms resulting from data processing are often communal in nature, given the interconnected nature of some data. Finally, ensuring truly informed consent for every individual has proven impractical.

We propose \textit{collective consent}, operationalized through \textit{consent assemblies}, as one alternative framework. We establish collective consent's theoretical foundations and use speculative design to envision consent assemblies leveraging deliberative mini-publics. We present two vignettes: i) replacing notice and consent, and ii) collecting consent for GenAI model training. Our paper employs future backcasting to identify the requirements for realizing collective consent and explores its potential applications in contexts where individual consent is infeasible.
\end{abstract}





\maketitle

\input{content/01_intro}
\input{content/02_background}
\input{content/03_framework}
\input{content/06_discussion}

\begin{acks}
This work has benefited from Dagstuhl Seminar 25112 ``PETs and AI: Privacy Washing and the Need for a PETs Evaluation Framework.'' We would also like to thank the anonymous reviewers, Abraham Mhaidli, and Gabriel Lima for their feedback which has helped improve this paper. Lin Kyi is funded by a Google PhD Fellowship and thanks them for their support. This research was conducted independently of Google.
\end{acks}



\bibliographystyle{ACM-Reference-Format}
\bibliography{sample-base}

\end{document}

%% file: content/01_intro.tex
\section{Introduction}
\label{sec:intro}
Gathering meaningful and informed consent is crucial for users to have more autonomy and control over their data. The primary way in which consent for data collection has been operationalized in practice by industry and regulators is through notice and consent. In notice and consent, users are first informed about how their data will be used (\textit{notice}), and then asked for their consent to agree to data processing (\textit{consent}), usually in the form of a consent notice. This model was inspired by the \textit{Fair Information Practices}~\cite{BerjonYasskin2025}, and has become the standard for collecting consent, being most notably adopted under the European Union under the General Data Protection Regulation (GDPR)~\cite{EUdataregulations2018} and ePrivacy Directive~\cite{EUDirective200258}. This model has also been adopted under the California Consumer Privacy Act (CCPA), which mandates that users have the right to know about data processing and opt out of data collection~\cite{CCPA}. Additionally, the Canadian government has laid out its own consent guidelines~\cite{can_consent}.

Despite its presence in the online space, notice and consent has often been criticized~\cite{utz2019informed, kyi2024doesn, nouwens2020dark, kyi2024reimagining}. The effects of the model are dubious; several research studies have shown that users are neither informed nor feel like they have control over their consent decisions~\cite{kyi2024doesn, utz2019informed, nouwens2020dark, grover2025data}, leading to feelings of digital resignation and privacy cynicism~\cite{grover2025data, van2024privacy, kyi2024doesn}. Much research and regulatory focus has been on consent notices. Still, these efforts do not address fundamental issues surrounding consent, which is that, under current implementations, consent is neither \textit{informed} nor \textit{meaningful}. 

Despite the shortcomings of the notice and consent model, a user's consent is important for technologies, and the companies, designers, and business practices responsible for them, to be ethical~\cite{singh2022consent, cintaqia2025stop}. Simply asking for a user's consent is an effective way of giving users control over their data~\cite{berjon2022consent}. 
Yet, current online data collection practices are usually extractive, ``wherein data is taken without meaningful consent and fair compensation for the producers and sources of that data''~\cite{sadowski2019data}. We need a new form of collecting consent that moves society away from the notice and consent model which can handle challenges to consent brought on by emerging technologies (e.g., generative AI), and instances in which a user's consent is difficult to obtain in the moment (e.g., public space surveillance, bystander privacy).

In light of recent calls to scale back privacy and AI regulations within the EU and UK in an effort to reduce regulatory restrictions and promote innovation~\cite{Hart_Europe_2025, hm_treasury}, we believe it is the ideal moment to \emph{reimagine consent}. We have an opportunity to put forth effective solutions that are human-centric and account for challenges that current consent mechanisms cannot address, while simplifying regulatory compliance for industry and users. 

Namely, we believe that consent would benefit more from a collective solution~\cite{BerjonYasskin2025}. A collective approach can address the communal implications of data collection and the collective harms we are facing from the same tech companies collecting and processing our data~\cite{wu2025design, zong2020individual}. Thus, a collective consent approach may provide users with more autonomy and control over their data and its outcomes, especially as it is unreasonable for users to be expected to be fully informed and make constant consent decisions regularly. 

In this paper, we present collective consent, imagined through consent assemblies, as a framework for responsible data collection. This is envisioned to not only replace notice and consent, but to handle consent in situations currently neglected by notice and consent, such as when data may be reused in the future (e.g., to train generative AI systems), and collect consent in other technology governance scenarios where collecting consent from individuals is difficult to obtain in the moment (e.g., smart homes). We argue that collective consent, compared to other individualized approaches, can protect users' privacy more effectively, reduce the burden of consent from individual users and industry, and can be implemented for a wider variety of data and technology governance scenarios.

We use speculative design, a methodology proposed by Dunne and Raby in their book, \textit{Speculative Everything}, to explore an alternate future for consent~\cite{dunne2024speculative}. We build upon ideas from privacy, political science, machine learning, and previous research on the limits of individual consent~\cite{walquist2025collective, zong2020individual, mozilla_consent, lovato2022limits, fairfield2015privacy} to ideate on a concrete and actionable collective consent framework that addresses the challenges of collecting data responsibly in a wide range of scenarios. We use future backcasting to envision what is needed to bring the idea into reality, including how to convince other stakeholders to adopt this alternative approach~\cite{servicedesigntools_futurebackcasting}.

In our operationalization, collective consent leverages deliberative mini-publics through what we call a \textit{``consent assembly''} to form collective consent decisions. In this process, a group of people affected by data processing is represented by a representative group (i.e., an assembly or panel) to make deliberate decisions about privacy and data collection on behalf of the broader population. This assembly will be presented with information regarding data processing from different stakeholders for a fuller picture, discuss whether they consent, and the conditions for their consent.

The purpose of this paper is to conceptualize how collective consent could function in practice. We lay the groundwork for a future research agenda focused on implementations and empirical studies of collective consent.


%% file: content/02_background.tex
\section{Theoretical Background}
\label{sec:background}
Here, we outline the theoretical background for our collective consent framework, drawing on ideas from previous consent systems and their alternatives, as well as concepts related to the collective governance of data.

\subsection{What Effective Consent Looks Like}
\label{sec:effective_consent}
Under the GDPR, the requirements for consent are that it needs to be \textit{freely given} (voluntary), \textit{specific} (allow for granular choices), \textit{informed} (provide clear information about data collection), and \textit{unambiguous} (affirmative act is needed to obtain consent)~\cite{EUdataregulations2018}. However, consent is more multifaceted than these requirements.

According to a report from the World Economic Forum, three characteristics of better consent alternatives are listed~\cite {wef_consent}. First, consent must allow for \textit{true user agency}, where users can negotiate or opt out of consent and have control over their data. Second, \textit{consent has to be participatory}, involving a diverse group of consent stakeholders to ideate solutions for consent together. Third, consent needs to involve a \textit{real choice} where users' consent is not permanent, but can be revoked and adjusted. 

Additionally, as advances in AI present new challenges to responsible data collection, sustainable consent mechanisms must be created, which work for both present and future technologies. Notice and consent is inadequate to deal with the challenges of GenAI~\cite{kyi2025governance}, and it is unreasonable to think about redesigning consent technologies for every new consent application. Instead, there should be more focus on the principles of effective consent and designing ecosystems that facilitate consent in these instances. 

\subsection{Why Notice and Consent is Ineffective}
\label{sec:background:ineffective}

\paragraph{User interface challenges.} The first criticism of notice and consent is that its user interface has poor usability~\cite{machuletz2019multiple, nouwens2020dark, kyi2024doesn}. Very few users pay close attention to consent notices, as they are often riddled with lengthy text and legal jargon~\cite{kyi2024doesn, santos2021cookie}. These lengthy notices, combined with deceptive designs made to trick users into consenting~\cite{kyi2023investigating, nouwens2020dark, utz2019informed}, and their constant presence in the online world, results in consent fatigue, where users become increasingly desensitized to the high frequency and complexity of requests for consent~\cite{kyi2024doesn, utz2019informed}.

\paragraph{Assuming every user can meaningfully consent and opt out.} Notice and consent assumes that everyone using these notices is capable of properly consenting because showing these notices to everyone forces every user to interact with them. However, not everyone using the internet and being exposed to notice and consent is capable of giving their consent. 

Under the CCPA, consent operates under an opt-out model where users can request to opt out of data collection, rather than opting in such as under the GDPR~\cite{CCPA}. However, children between the ages of 13 and 16 must provide explicit consent for data collection, and for children under 13 years old, the consent has to come from their parent or guardian~\cite{CCPA}. Though there is recognition that children deserve to consent and have different consent considerations from adults, the collection of a child's consent is likely to be less practical in reality~\cite{wang2024chaitok}. More broadly, asking for consent from vulnerable user groups will be less straightforward in practice. For instance, those who lack computer skills, those who are intellectually disabled, or those who are interacting with consent notices in a language they do not understand may not be able to make a properly informed decision~\cite{wang2024chaitok, consent_disability, ding2024up, consent_language}. 


Notice and consent is also unable to account for the complex relationships between data, leaving some users unable to properly consent despite their information being revealed. \citet{lovato2022limits} noted that consent notices and individual consent are inadequate for social networks because data is distributed across multiple users, making it difficult for all affected users to properly consent, and proposed a platform-specific distributed consent system. In a study about data donation for online messaging, \citet{walquist2025collective} recommended that collective consent should be a consideration to include in data donation platforms to improve privacy for online interactions involving several people.


\paragraph{Notice and consent applies for a narrow range of data collection practices.} 
Consent notices tend to only cover purposes related to web tracking~\cite{kyi2024doesn, machuletz2019multiple}, but these are not the only types of data being collected. Online data collection is a complex ecosystem, and consent notices only work for consent \textit{in the moment}, not for \textit{retroactive consent} (i.e., consent for data that has already been collected), \textit{future uses} of this data (i.e., consent to reuse this data in the future), nor for other instances of data collection where asking for consent in the moment is impossible. 

In the context of retroactive consent, \citet{kyi2025turning} found that industry practitioners often ended up relying on other GDPR legal bases to collect data since consent could not be applied to these cases. 
In addition, consent notices are ineffective for future uses of consent, where data is reused for other contexts, such as in the case of Generative AI (GenAI)~\cite{asthana2024know, kyi2025governance, pasquale2024consent, biega2023data}, where already-collected data is re-used to train AI models that can generate content. The inappropriate reuse of data to train GenAI systems has led to concerns over the lack of consent obtained to train GenAI systems~\cite{kyi2025governance, pasquale2024consent}. It is difficult to establish the provenance of data in complex AI systems~\cite{wang2015big, longpre2024bridging}. Thus, asking for consent individually is difficult, and we encourage exploring collective options.

Furthermore, data collection often occurs in contexts where obtaining individual consent is impractical or impossible. Examples include surveillance in public spaces~\cite{zhangmore}, smart homes~\cite{zeng2017end}, and privacy intrusions stemming from social networks and interconnected data ecosystems~\cite{lovato2022limits}. This represents a fundamental limitation of the notice and consent framework: we cannot feasibly notify and obtain consent from all affected individuals for every instance of data collection, particularly beyond traditional web tracking scenarios.

\subsection{Alternatives to Notice and Consent (and the Issues they Present)}
\label{sec:background:alternatives}
Due to the challenges that notice and consent present, others have introduced solutions to reduce the number of consent notices users need to interact with. However, these new solutions can either exacerbate pre-existing challenges, introduce new challenges to users, and do not address fundamental shortcomings of current implementations of consent addressed in Section \ref{sec:background:ineffective}. 

\subsubsection{Browser-Level Consent (Global Privacy Control)}
One commonly-mentioned alternative to consent notices is browser-level consent, the most common one being Global Privacy Control (GPC).\footnote{\url{https://globalprivacycontrol.org/}} GPC was created in response to the California Consumer Protection Act (CCPA), allowing users to consent to cookies just once using the browser, rather than individually for each website~\cite{GPC}. GPC is available only on some browsers, such as Firefox, Duck Duck Go, Brave, or additionally as a browser extension~\cite{GPC}. 

While Do Not Track was unable to reach mass popularity, GPC (and browser-level consent) is generally seen as the future of consent, being usable and requiring fewer consent decisions from users~\cite{zimmeck2023usability}. However, there are still major issues with GPC and similar technologies, making it an inadequate alternative to consent. 

First, GPC is still at the mercy of powerful tech companies, who can choose to implement this privacy control or not. Notably, GPC is not offered with popular web browsers such as Google Chrome or Apple Safari~\cite{google_GPC}. Second, browser-level consent does not allow users to make granular decisions, such as sharing only certain types of data for certain websites. Previous research has shown that users do not perceive data collection similarly across all sites or domains, often preferring to share data they perceive as being relevant to accomplish a specific task~\cite{sharma2024m, sleeper2016sharing}. Lastly, the onus of understanding and making an informed consent decision is still on the user with browser-level consent. This is an issue because if the user is unable to make informed decisions, such as if they are a child, lack technology literacy, or have other factors making them unable to be informed and consent meaningfully, browser-level consent does not solve the issue of collecting informed and meaningful consent.

\subsubsection{Reject-All Browser Extensions}
In response to the plethora of consent notices, browser extensions enabling users to reject all cookies, such as Do Not Track (DNT), have been proposed. The browser extension sends a signal to all the websites a user visits, automatically declining tracking~\cite{DNT}. 

Despite these technologies being in the interest of users, DNT and similar browser extensions were unable to reach critical mass because of a ``lackluster adoption by browsers and a lack of mechanisms to enforce the user's preference''~\cite{DNT}. DNT was also unsuccessful because web standards, such as the W3C, did not adopt it as a standard, and web browsers had their own tools to block and limit tracking, making DNT ineffectual~\cite{DNT}.

On top of these issues, reject-all browser extensions present other issues for users. First, reject-all browser extensions require know-how to navigate. DNT is largely forgotten, but there are many other browser extensions that now exist, with different functions and abilities. Users will need to know which browser extensions are more effective, how to use them, and trust whether they are actually working to reject all tracking. Second, data refusal comes with tradeoffs, such as reduced accuracy~\cite{vincent2019data}, and users' intentions may not always be to reject all tracking, depending on the purpose or website~\cite{kyi2024doesn}. Third, reject-all browser extensions can present fairness issues; those who use these extensions will likely differ from those who do not, resulting in an under- and over-representation of data collection for certain demographics.

\subsubsection{Predictive Consent}
Another idea to replace consent notices is to use predictive consent or AI agents, where predictions about a user's privacy preferences are made based on data about the user~\cite{muhlhoff2023predictive}. Predictive consent and predictive privacy, in general, present several issues for users. 

First, AI agents can introduce user concerns around trustworthiness (i.e., low accuracy) of the agent, can disempower users, and there is the risk that AI agents may be used maliciously~\cite{chan2024visibility}. Second, predictive privacy requires a lot of user data to train these models, which can present additional data privacy issues to the user~\cite{muhlhoff2023predictive}. A previous study has shown that users' digital footprints can accurately predict certain demographic factors~\cite{hinds2018demographic}, showing that privacy is a reasonable concern with this method. Third, predictive privacy presents an additional informational asymmetry between users and online platforms~\cite{muhlhoff2023predictive}; users do not know what these predictions know about them, nor the kind of data that was used to train these predictions.




\subsection{Collective Governance of Data and Privacy}
\label{sec:background:collective-governance}

Many works in machine learning and related areas have demonstrated that collective action can help promote user rights~\cite {hardt2023algorithmic, vincent2021data, karan2025algorithmic}. Computing systems rely on the data from many to create and optimize their systems. Previous work by \citet{vincent2021data} has shown that \textit{data leverage,} referring to users conducting data strikes, data poisoning, and conscious data contribution, can harm computing systems and give users more leverage over powerful tech companies. As noted by \citet{zong2020individual}, individual consent does not provide users with true autonomy, and the refusal of consent from one individual does not equate to broader and systemic changes to data collection practices. Instead of individual consent, Zong argues that ``Collective action can help people address the autonomy problem, manage data over time, and replace individual disempowerment with collective power''~\cite{zong2020individual}. 

Thus, we need to consider collective forms of governing user data to ensure its responsible usage by companies, such as by implementing collective consent. Although privacy and data are commonly thought of as one's own commodity to create and control, many works argue that privacy is a public and collective good. 

It is arguably more effective to govern data as a public or communal commodity rather than an individual one for several reasons. First, not every user has the ability or interest to act in ways that protect their privacy, thereby revealing information about other users~\cite{fairfield2015privacy, wef_consent}. Second, users are largely unaware of how the data collected today may impact them in the future, especially as the impacts of individual instances of data collection seem small until combined~\cite{wu2025design}. Third, governing data as a collective commodity can help achieve optimal privacy for the masses~\cite{saetra2020privacy}. \citet{saetra2020privacy} notes that governments need to have a role in collective privacy management, as ``privacy has value beyond what most individuals can foresee and the government has a duty to prevent new technologies from disrupting the provision of this good''~\cite{saetra2020privacy}. Fourth, negotiating privacy preferences between various participants can be a feasible task; almost all (97.5\%) scenarios they presented to participants reached an agreement in just over three minutes in a study by \citet{zhou2024bring}. This shows that it is possible for users to collaborate on privacy management.

Collective data governance, and by extension collective consent, can help strengthen privacy protections for the \textit{majority} of users, as noted by previous works calling for collective approaches to privacy~\cite{fairfield2015privacy, walquist2025collective, rudolph2018users, dupree2016privacy, zong2020individual}. Since issues with consent are largely social, we believe that sociotechnical solutions, such as consent assemblies, are more promising than past technological approaches. In this paper, we conceptualize how collective consent can be implemented, bridging the gap between the legal requirements for consent and ensuring this consent is meaningfully informed, and comprehensive to account for data collection not covered by notice and consent.



%% file: content/03_framework.tex
\section{Collective Consent Through Deliberative Mini-Publics}
\label{sec:framework}

Building on the model of \emph{deliberative mini-publics}, which are increasingly used in the policy-making domain~\cite{oecd2020innovative}, we propose \emph{consent assemblies}. These are bodies composed of randomly selected participants who arrive at a position on consent through a process of joint deliberation.
We develop this concept of consent assemblies throughout the remainder of the paper.


\subsection{Background on Deliberative Mini-Publics}
\label{sec:background:sortitions}
Deliberative mini-publics are a democratic process in which members of a wider population are randomly selected and together weigh in on a policy question on behalf of this population~\cite{lowbeer2022democratic}. The assembly is a ``mini'' version of the jurisdiction (such as the nation where it takes place). Assemblies typically have a similar demographic make-up to the wider population~\cite{de_bundestag}, along dimensions such as gender, racial, socioeconomic, educational, or geographic diversity, ensuring that the deliberating body reflects the perspectives and lived experiences of the broader community it represents. The panel of randomly selected members undergoes an extensive process of learning and discussions about the topic before formulating joint recommendations to policymakers~\cite{oecd2020innovative}. 

The idea of decision making by random citizens goes back to Athenian democracy~\cite{Hansen91} and has been regaining traction since the 2010s~\cite{democracynext2023assembly,oecd_innovative_2025}.
High-profile examples include mini-publics on same-sex marriage, abortion, and gender equality in Ireland~\cite{Courant21}, and the French citizens' convention on climate change~\cite{GAA+22}.
Recently, deliberative mini-publics have been institutionalized as permanent parts of governance in a number of cities, including Paris, Brussels, and Copenhagen~\cite{fide_permanent_assemblies}. Typically, deliberative mini-publics see high rates of agreement. In a study by \citet{lage2023citizens}, 75\% of proposed policies have been found to have an agreement rate above 90\%, and 90\% of proposed policies have an agreement rate above 70\%.

An attractive property of deliberative mini-publics is that its members engage with the policy question by seeking out information, reflecting on the issue personally, and discussing it with others\,---\, a much deeper engagement than is practical for all voters in an election~\cite{fishkin2018democracy}.
Another advantage is that by selecting members as a representative sample of the population, all subgroups are represented in proportion to their population share, in contrast to known differences in turnout between demographic groups~\cite{LN14}.
Empirical research on deliberative mini-publics has found that mini-publics engage in high-quality deliberation, can overcome polarization, and are not easily manipulated by elite framing~\cite{DBC+19}.

There is precedent for using deliberative mini-publics in the context of tech policy~\cite{oecd_innovative_2025}.  
For instance, \emph{alignment assemblies} organized in collaboration with Anthropic and OpenAI 
have discussed what the public wants out of AI, and how to align AI to better support humans~\cite{cip2024alignment}.
Another recent example is a 2024 deliberative mini-public in Germany on disinformation~\cite{forum_gegen_fakes}, in which members furthermore interacted with input from thousands of online participants.

Deliberative mini-publics go by a variety of names (``citizens' assemblies'', ``citizens' panels'', ``deliberative polls'', sortitions, to name a few), which carry different connotations about the questions being discussed, the number of panel members, and the duration of the process.
At the larger end, \emph{citizens' assemblies} address weighty and complex political questions, on which a panel of about 100 participants deliberates in depth over about 20 days~\cite{oecd2020innovative}.
At the other end of the range, \emph{citizens' councils} often involve just 15 panel members for one or two days of deliberation.
Though we term our mini-publics \emph{consent assemblies}, we do not aim to make a one-size-fits-all decision about the size and depth of the process, which should be commensurate with the complexity of the question and with resource availability constraints. 

Broadly, the steps of a deliberative mini-public can be broken up into the following~\cite{democracynext2023assembly, ie_selection, de_bundestag}:
\begin{enumerate}
    \item \textit{Inviting and Selecting Members:} A random sample of the population is invited and selected to participate in a citizens' assembly.
    \item \textit{Learning and Listening Phase:} The assembly members have in-depth learning sessions about the topic of discussion.
    \item \textit{Deliberation Phase:} The assembly members discuss their opinions and deliberate on the potential outcome(s).
    \item \textit{Outcomes:} The assembly makes recommendations to policy-makers and other involved parties about the particular outcome of their discussion.
\end{enumerate}



\subsubsection{Inviting and Selecting Members}
Invitations are sent to a random sample of approximately 10,000 to 30,000 people in the given population~\cite{democracynext2023assembly}. These invitations are sent by mail, email, phone, and other ways to reach these potential members. Those who opt in to participating are put in a random lottery to select the members of the deliberative mini-public~\cite{democracynext2023assembly}. The selected individuals in the lottery undergo another lottery, this time stratifying them by demographics (for instance, age, gender, socioeconomic status) to eventually have an assembly that is representative of the population, including non-citizens of a jurisdiction~\cite{democracynext2023assembly, ie_selection}. 

Approximately 30 to 200 participants from a given population are randomly selected~\cite{de_bundestag} through a process of ``sortition'' or ``civic lottery''~\cite{uk_CA}. Participants are stratified based on demographics to ensure a representative group is selected. The involvement of politicians or anyone with personal or professional interest in the matter cannot be members, as their opinions will be biased~\cite{participedia}. The selected members of an assembly are supposed to be representative of the population in terms of demographics, experiences, and backgrounds~\cite{participedia, flanigan2021faira}. Thus, deliberative mini-publics will typically include a diverse range of members to maintain representativeness~\cite {ie_selection}.

Computer science research on citizens’ assemblies has mainly focused on algorithms for sampling assembly members~\cite{ebadian2022sortition, flanigan2020neutralizing, flanigan2021faira, flanigan2021fairb,GMS+25}. \citet{flanigan2021faira} developed fair algorithms for selecting panels for deliberative mini-publics, and one of their proposed algorithms has been implemented in at least 40 citizens' assemblies globally. To select the panel, they must find members who are representative of the greater population, and ensure that there is an equal likelihood of everyone being selected to join~\cite{flanigan2021faira}. Oftentimes, it is believed that uniform selection, in which a random, uniform panel of people is selected, is considered the ideal way to construct a sortition~\cite{engelstad1989assignment}. In panel participation, it is still common for some people to drop out. To mitigate the impacts that dropping out has on demographic fairness, \citet{flanigan2020neutralizing} developed an algorithm to predict the likelihood of different panel volunteers meeting certain demographic criteria, thereby preventing over- and under-representation of specific demographics.

\subsubsection{Learning and Listening Phase}
As the members of the deliberative mini-public have different levels of knowledge and experiences with the task to be discussed, they need to be prepared, which involves extensive learning about the task~\cite{participedia}. Depending on the complexity of the matter, this can range from as short as one day to a couple of weeks or a few months~\cite{participedia, oecd2020innovative}. Learning sessions can involve a variety of tools to facilitate learning Q\&A sessions with different stakeholders, panels with experts on the topic, and smaller group discussions to help improve understanding of the material~\cite{participedia}. The listening phase allows for discussions, questions, and comments from assembly members. 

\subsubsection{Deliberation Phase}
Members of the deliberative mini-public will debate and critically engage with the information they were previously presented with. In these discussions, a neutral facilitator will usually moderate the proceedings. Members can exchange experiences, opinions, and arguments, and only then will organizers mention that the goal is to reach a decision~\cite{participedia}. 

In the deliberation phase, the design of groups is crucial to ensure that everyone can voice their opinions and is exposed to alternative ideas. Typically, members deliberate through a mix of small-group and larger plenary sessions. Since not everyone will be equally comfortable with sharing their opinions with the larger group, smaller group discussions are very important~\cite{participedia}. It is also important to adjust the seating arrangements to facilitate exposure to diverse ideas and perspectives. Specifically, organizers of a citizens' assembly will design a seating plan to allow for distribution of attitudes and demographics~\cite{social_care}.

\subsubsection{Outcomes}
Citizens' assemblies typically make some recommendations or decisions which are then presented to policy-makers, and perhaps the general public~\cite{InstituteforGovernment_CitizensAssemblies}. Oftentimes, policy recommendations are only shared with policy-makers for decisions with over 75\% consensus among assembly members~\cite{Abels2022NextLevel}.

Some jurisdictions, such as Ireland, have a formal process where citizens' assembly recommendations are reviewed by the relevant parliamentary committee, which considers and presents the report to the government~\cite{InstituteforGovernment_CitizensAssemblies}. Citizens' assembly decisions are not always binding. Instead, these decisions are shared with governments and policy-makers~\cite{ferejohn2008citizens} who may set up an implementation timeline and need to provide reasons in case they do not accept some of the recommendations from the assembly~\cite{InstituteforGovernment_CitizensAssemblies}. Typically, summaries of the assembly process will be shared publicly to maintain transparency for the public~\cite{boell}.

\section{Collective Consent Assemblies}
With this background on deliberative mini-publics, this section explores how the process can be adapted for user consent in digital systems -- in the form of \emph{collective consent assemblies}.

\label{sec:consent_assembly}
\subsection{Methodology}
Given the scope of the problem, to ensure adequate expertise coverage, our research team included experts in HCI, privacy, algorithmic fairness, information systems, operations research and resource allocation, and internet and data governance. Team members also have prior experience as practitioners in the technology governance and policy space, as well as collaborating with nonprofits specializing in deliberative democracy. 

We employed speculative design to conceptualize collective consent. Speculative design is a method for futuring (ideating about the future), and has been applied in other HCI projects, either as a theoretical contribution for envisioning design futures~\cite{harrington2022all}, or as a method for co-design with participants~\cite{tao2025housing, cho2024reinforcing}. 

Speculative design is a way of reimagining or inventing designs to meet the true needs of users, and is used to address larger societal issues that are beyond the limits of traditional design~\cite{dunne2024speculative}. As current-day implementations are unable to meet the demands for truly informed and meaningful consent, we employed a speculative design methodology to think of a radical alternative to notice and consent. This approach suits collective consent because the process will require fundamental systemic changes beyond surface design or regulatory modifications. Speculative design allows us to examine broad, complex problems by considering the larger ecosystem of factors involved.


We structured the ideation process into several phases:

\paragraph{Translating Deliberative Mini-Public Frameworks}
We grounded our approach in several established deliberative mini-public guidelines \citep{democracynext2023assembly, participedia, de_bundestag}. Members of the team regularly met to engage in speculative design discussions. The team imagined scenarios for each step of the consent assembly process based on the deliberative mini-public process, then discussed how each step could be materialized for different scenarios. At this stage, our team's design discussions focused on how to adapt these participatory democratic processes to the context of consent decision-making. 

\paragraph{Vignettes}
In the second stage, we created specific case studies. We developed hypothetical, realistic vignettes to demonstrate the practical implementations of our proposal for collective consent assemblies and to include additional considerations for these practical implementations. Through several regular meetings within the research team, we brainstormed several potential applications for collective consent assemblies to form the vignettes. 

The process of ideating about collective consent through the vignettes was iterative. The initial process for conducting a deliberative mini-public helped further describe aspects of implementing collective consent using the vignettes, while the vignettes also contributed to the development of the collective consent assembly framework. We updated our proposal for collective consent assemblies based on the detailed considerations presented by the vignettes, repeating the process until the research team converged on an agreement that the proposed framework was specific enough to envision its practical implementations, while being broad enough to apply to other consent situations. The first author then collated these ideas to create the vignette narratives.

After identifying a superset of potential applications, we decided to proceed with two vignettes, \textit{Notice and Consent Replacement} and \textit{Future Reuse of Consent and Data for GenAI Model Training}. The first vignette presents a common problem that society is currently facing, while the latter presents a newly emerging issue. With these two vignettes, we demonstrate the broad applicability of collective consent assemblies in addressing the challenges of responsible data governance.

Finally, we sent our vignettes, including the scenarios and steps of the consent assembly, to an external expert on deliberative mini-publics for review. We have asked the expert to evaluate whether the vignettes are realistic to how deliberative mini-publics function in practice, and specific enough so that a real-world assembly could be implemented using the provided description. We revised our vignettes based on the expert feedback.

\paragraph{Backcasting.}
Using our collective consent framework and vignettes as a starting point, we applied future backcasting to identify the present-day changes necessary to achieve the vision of collective consent in Section~\ref{sec:backcasting}. \textit{Future backcasting} is one method for futuring~\cite{servicedesigntools_futurebackcasting, vergragt2011backcasting, bendor2021looking, Gold2023Backcasting}. It is about figuring out what \textit{should} happen \emph{right now} to make a future possible~\cite{vergragt2011backcasting, Gold2023Backcasting}. 

Backcasting recognizes the systemic and society-wide challenges that exist in the way of futuring (in our case, implementing collective consent assemblies), and recognizes that transformations are needed in the here and now to eventually achieve a desirable future~\cite{vergragt2011backcasting}. We identified four areas of transformation together with their key stakeholders (policymakers and regulators, businesses, users, society), and systematically examined the required actionable next steps.


\begin{figure*}
  \includegraphics[width=0.6\linewidth]{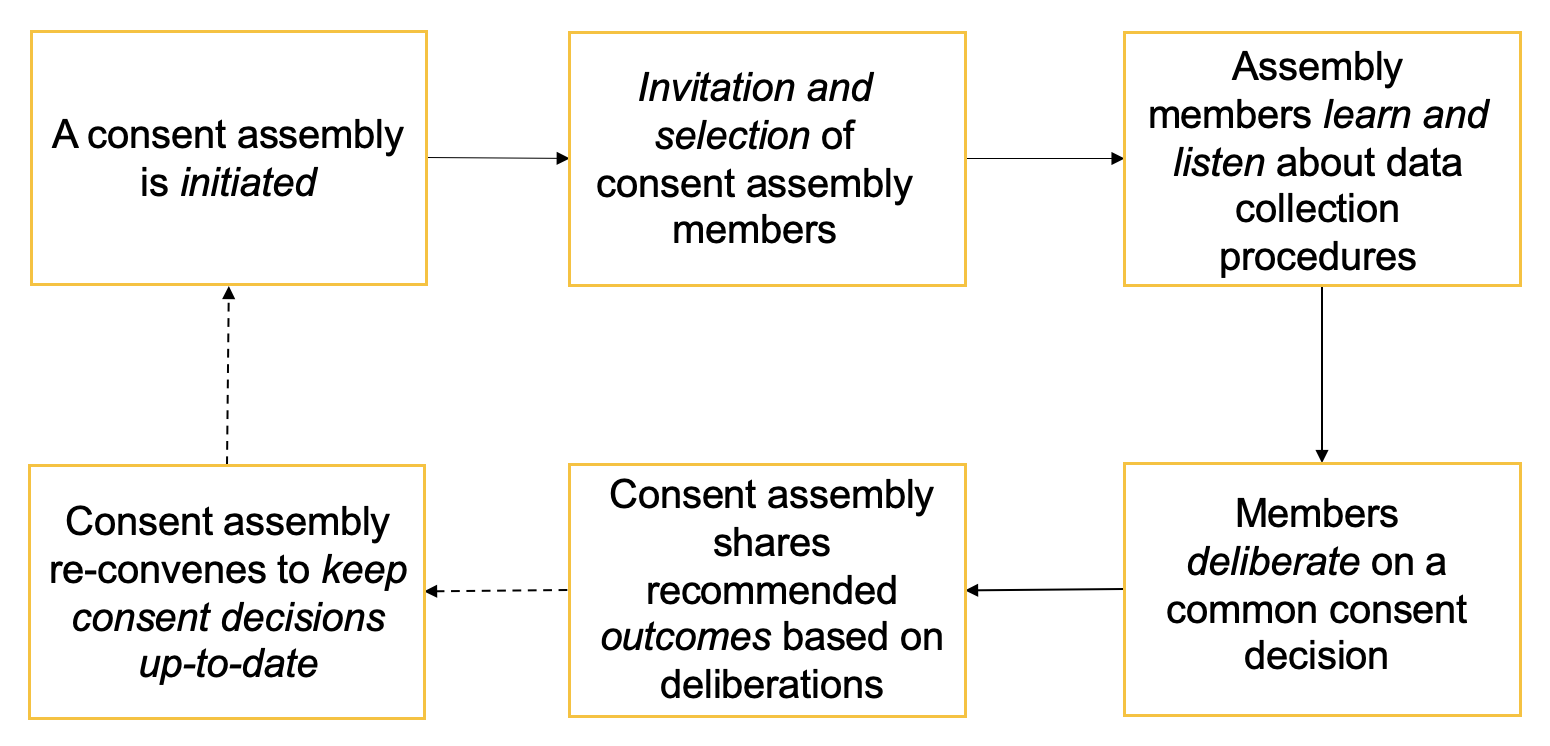}
  \caption{The proposed steps of a consent assembly. The solid lines indicate steps that \textit{could} happen in every consent assembly, and the dotted lines indicate steps that \textit{might potentially} happen depending on whether the situation requires it or not.}
  \label{fig:consent_diagram}
\end{figure*}

\subsection{Initiating a Consent Assembly}
A request for consent will be submitted, either by: i) regulators (for wider-scale consent deliberations) or ii) companies (for deliberations about a given company's practices). This request is envisioned to be submitted in writing by a party seeking a collective consent decision to proceed with data processing. The request should outline the proposed data processing, its rationale, and the potential impact it may have. 

\textbf{Who runs consent assemblies.} It is important that consent assemblies are organized and operated without bias to promote the collective's interests. Privacy regulations are already overseen by dedicated entities such as national data protection authorities (in the EU and Brazil), and the Office of the Attorney General for the CCPA, who could potentially oversee collective consent assemblies. There could also be participation from privacy NGOs or NGOs that run deliberative mini-publics in the consent assembly process. Additionally, there is the possibility that new agencies could be established to facilitate the process of running and organizing consent assemblies. 

Industry organizations that are interested in data collection will also have a chance to participate in consent assemblies. Industry participants could be allowed to have one or more representatives attend the assembly to present their perspective on the deliberated data processing practice.

\subsection{Inviting and Selecting Consent Assembly Members}
In addition to stratifying for typical demographic factors, such as those found in a deliberative mini-public, a consent assembly may need to ensure that various privacy preferences are represented in the assembly and avoid privacy biases. Privacy preferences generally refer to a member's attitudes and/or behaviors towards privacy and data sharing. Privacy preferences may need to be requested in the initial invitation, or whenever people from a jurisdiction enter their demographic information, such as during a census or when expressing interest in participating in consent assemblies. 

\textbf{Who participates in consent assemblies.} For users to feel a consent assembly is representative, it is important that the assemblies are reflective of as much of the population's diversity as possible. 
Some representation metrics, such as a person's demographics, are easier to estimate and account for when assembling consent assemblies. Other metrics, such as privacy preferences, may be more difficult to measure. Given that consent involves one's personal data, consent assemblies need to be careful in how users are represented. 

Deliberative mini-publics tend to see high rates of approval by the public~\cite{lage2023citizens}. However, aggregating all users and disregarding specific circumstances can cause privacy harms, especially as consent assemblies relate to privacy and personal data.
People may have different considerations not only due to their demographics (e.g., children) but also experiences (e.g., those without technical knowledge). It is therefore important not to silence these groups.

\textbf{Stratifying assemblies by demographics, privacy opinions, and other factors.} To find a balance between representativeness and preventing collective interests from eroding the interests of specific groups, consent assemblies may need to stratify based on demographics, privacy opinions, and other factors. For instance, not all users care equally about privacy, which may result in those who value it being dissatisfied with the collective's decisions. 

Different demographics may also have specific concerns related to consent and privacy. For instance, historically marginalized groups may also have different consent concerns and vulnerabilities, which would require them to make their own collective decisions that differ from those of the general population. Based on the sub-group's decisions, a collective consent decision can then be applied broadly to the wider population that this sub-group represents.

To operationalize representativeness, entities that oversee consent assemblies may consider distributing surveys to the general public, or relying on census data before initiating a consent assembly. These preliminary data collection efforts can help identify relevant stratification criteria and ensure that assembly composition reflects the diversity of affected populations. Surveys might assess not only demographic characteristics but also privacy attitudes, digital literacy levels, past experiences with data misuse, and specific vulnerabilities that could inform meaningful subgroup formation.

\textbf{Making consent assemblies accessible to participate in.} Improving access is crucial for consent assemblies to be representative. To improve accessibility, it is worth considering whether consent assemblies might be able to provide travel funding, childcare, or compensate participants for taking time off, similar to how jury participants for court cases are compensated by the government. 

To reduce costs and avoid location-based constraints for participation, others have experimented with online discourse and platforms for online deliberation, sometimes with an AI-based facilitator, showing promising results~\cite{fishkin2019deliberative, argyle2023leveraging}. Having remote participation also allows those with mobility issues to participate, along with the ability to add live translations for individuals who struggle with the majority language or reside in jurisdictions with multiple languages.

\subsection{Learning and Listening Phase}
It is important to involve a variety of stakeholders, such as those from industry, regulators, privacy lawyers, NGOs, etc., to expose members to a variety of views and allow for exchange of dialogue between members and these stakeholders. As members may have pre-conceived notions of what different purposes and companies do with user data~\cite{kyi2024doesn, chanchary2015user}, it is important not to have these biases impact collective consent decisions. Instead, the focus of the \textit{Learning} and \textit{Listening Phase} should be on educating members about the costs and benefits of data collection for various purposes.

\textbf{How to present complex information to laypeople.} Data processing is not the priority for all users, nor is it something most users can easily comprehend without breaking these concepts down into understandable ideas~\cite{kyi2024doesn}. This complex information must be presented clearly and engagingly. 

We recommend that consent assemblies use teaching tools and techniques tailored to the specific groups and situation. Consent assemblies could consult prior works from psychology, which have recommended teaching techniques. For example, \textit{scaffolding} could be applied to help people learn new ideas while adapting to their needs and progress~\cite{vygotsky1978mind} such as by making analogies to familiar concepts, providing concrete examples before abstract principles, and gradually reducing support as understanding develops. Considering users' \textit{cognitive load} in the \textit{Learning and Listening Phase} could be applied by limiting the amount of new information presented at once, breaking sessions into manageable segments, and using visual aids to complement verbal explanations to make learning more attainable~\cite{sweller1988cognitive}.

\subsection{Deliberation Phase}
The \textit{Deliberation Phase} should allow members to deliberate on consent decisions. Ideally, members should also be able to \textit{negotiate} their consent terms and conditions with organizations collecting data, and deliberate on conditions for consent. Consent assemblies should allow for refusing consent to data collection if members agree that data collection in a certain situation is more harmful than beneficial.

Deliberative mini-publics can work for complex topics. Previous studies on citizens' assemblies have shown that members, when working together, can see through framing and manipulation by powerful entities to get members to decide in a way that aligns with what a company wants~\cite{niemeyer2011emancipatory, druckman2003framing}.

\textbf{Finding agreement in a collective decision-making process.} Previous work by \citet{zhou2024bring} has found that in their study of collective privacy management for IoT devices, participants came to agreement almost all the time, and compromise is an important part of the collective privacy management process. 

Sometimes, participants in a consent assembly may disagree on their decisions, which may cause privacy-related harms to others. In the \textit{Deliberation} phase, it is important for facilitators to encourage negotiation and discussion between consent assembly participants to ensure they reach a conclusion that most are satisfied with.

\textbf{Consent assemblies deliberate by purpose.}
We recommend that consent assemblies deliberate on consent by \textit{data collection purpose} (e.g., advertising attribution, bystander privacy for satellite imagery), and then make more specific consent decisions based on the kind of entity they are sharing data with (e.g., healthcare, e-commerce, social media). 

We believe that deliberating by purpose, and then making more granular consent decisions, such as those based on service type and/or trust in different companies, will make collective consent more feasible to implement because organizing a separate consent assembly for each company will be difficult, and many more assemblies will need to be organized compared to deliberating by data collection purpose. 

This purpose-based approach aligns with the principle of contextual integrity, which theorizes that privacy violations occur when information flows in ways that violate context-specific norms~\cite{nissenbaum2004privacy}. By focusing deliberation on purpose rather than entities, consent assemblies can better evaluate whether data practices respect the informational norms appropriate for different data-sharing contexts. 

\textbf{Consent is multi-faceted: Purpose, organization, and data types may influence consent decisions.} Though consent assemblies are theorized to be split up by purpose, we foresee that the kind of organization collecting the data (e.g., social media, news, e-commerce, charity, etc.) and the kind of data (e.g., personally identifiable data, behavioral data, etc.) will also impact consent decisions. Consent assemblies take into account the multifaceted nature of consent, allowing members to make more granular decisions based on these dimensions. 

\subsection{Outcomes}
In this stage, recommendations from the consent assemblies are shared with those involved in data processing, such as regulators, industry, and others with a stake in data collection. As consenting to data collection has privacy implications, in line with traditional deliberative mini-publics, it is suggested to only share policy recommendations where the assembly reached at least 75\% consensus~\cite{Abels2022NextLevel}. To make collective consent a concept that works for the majority, there should be a high level of agreement for consent (or opting out).

Ideally, regulators and industry would actively implement collective consent recommendations into practice and enforcement activities could focus on whether companies are complying with these consent recommendations. To ensure transparency, it is essential for regulators and industry to clearly explain when they are not following an assembly's recommendations, provide a justification, and consider the possibility of re-running a consent assembly in the event of a disagreement.

\textbf{Types of outcomes for consent assemblies.} Collective consent assemblies may come up with the following recommendations for consent:

\begin{itemize}
    \item \textit{Accept Consent:} The consent assembly agrees with data processing for this purpose, without any conditions for consent.
    \item \textit{Conditional Consent:} The consent assembly will only accept data processing for this purpose, under certain conditions. These conditions may include the requirement that only certain companies can collect this data, the data can be retained for a specific period, or any other condition.
    \item \textit{Reject Consent:} The consent assembly disagrees with data processing for this purpose, and there will need to be an updated proposal for data processing, and another consent assembly will convene to discuss. 
\end{itemize}

\subsection{Keeping Consent Decisions Updated}
Though not a step in the deliberative mini-public process, consent needs to be an ongoing process. This is especially important as technologies and regulations are constantly changing~\cite{kyi2025governance}. As such, we recommend that if there are any changes to technologies and regulations in a society, the process of reconvening a consent assembly needs to occur to keep the dialogue going. 

\textbf{Who participates in consent updates.} To make consent assemblies more feasible and avoid potential influence from a prior consent assembly, it is recommended that a new group of members be selected. However, the same demographic groups selected for the original assembly, plus any new and relevant demographic groups, could be selected for specific consent assemblies.

\section{Vignettes}
In this section, we connect the consent assembly structure proposed in Section \ref{sec:consent_assembly} with two vignettes to demonstrate how this could function in practice. These vignettes are not an exact recipe for how consent assemblies need to be implemented in the future, but rather an imagination of how collective consent could look like and how to think through various consent- and implementation-related considerations.

\subsection{Vignette 1: Notice and Consent Replacement}
The national data protection authority (DPA), overwhelmed with complaints from users about non-compliant consent notices, has decided to commission a consent assembly across the nation to decide on how to set up a default consent decision for users across websites for UX improvement purposes. Their goal is to understand which uses of UX improvement purposes are permissible, what data users are willing to share for these purposes, and see if there are any conditions for consent (such as limiting consent depending on the industry, company, or type of data that is being collected).

Many companies and organizations in the same jurisdiction, subject to the GDPR, want to collect data for user experience (UX) improvements, a purpose that is commonly found in consent notices. In exchange for collecting data about users' habits and behavior patterns on a website, including clickstream data, A/B testing performance metrics, and user interactions, the company can optimize the website along several key metrics, such as usability, accessibility, and performance. 

Many analytics teams across these organizations and companies have noticed that most users click ``accept all cookies'' or increasingly, ``reject all cookies'' without engaging very much with the banner, even though the second page of the consent notice explains what these purposes are and what they are doing with user data. Rejecting all cookies is frustrating for these organizations and companies, as they believe they are collecting data for a meaningful purpose, and do not want to lose important user data. However, consent notices do not lend themselves to nuanced decisions.

In this vignette, we explain how consent assemblies can be applied to the case of UX improvement purposes, as a way towards potentially replacing notice and consent in the future.

\textbf{Inviting and Selecting Consent Assembly Members:} As many websites typically engage both \textit{passive users} (i.e., those not signed into an account and just browsing online) and \textit{active users} (i.e., those with an account and signed in), the consent assembly should account for how experiences may differ between these groups. 

UX improvement purposes can have implications for accessibility. Therefore, the data protection authority determined that a separate consent assembly would be held for those who use online accessibility features.

To account for both passive and active users on these platforms, the data protection authority sent a survey to thousands of users asking them if they have accounts with these websites to be discussed, or browsed them without an account. Then, a random selection of those who answered ``yes'' and ``no'' were invited to the consent assembly to deliberate on this topic. The survey also asked whether participants used accessibility features, and those who selected ``yes'' were also invited to participate in the assembly for individuals with accessibility needs.

\textbf{Learning and Listening:} In this phase, members of the consent assembly have the chance to understand the potential benefits and harms of processing user data for UX improvements. Companies that participate have the opportunity to explain to users why they collect certain types of data and what they do with it. The data protection authority running this consent assembly determined that experts in other domains, such as law and engineering, should be included in this phase to provide another perspective on user data collection that may counter industry perspectives. 

Some members of the consent assembly previously believed that UX improvement purposes were the same as advertising purposes, and used to covertly deliver more advertising to users. Through the \textit{Learning and Listening Phase}, companies and other experts were able to clarify these misconceptions with assembly participants, explaining how the industry collects user data, how much data is needed, what the purpose of ``UX improvement'' entails, and the legal aspects surrounding the appropriate use. In turn, assembly members were able to ask questions of these stakeholders, challenging the importance of the amount of data being collected and the purpose of its collection.

\textbf{Deliberation:} Members of the consent assembly deliberated about whether they are convinced that sharing data for UX improvement purposes is actually of benefit to users, or mainly the company. 


Of particular importance were concerns about nudging, deceptive patterns, and UX-driven manipulation, all of which are drivers of interest that companies have in collecting data for ``UX improvement'' purposes. The deliberation focused on the limitations that will have to be placed on experiments that the company can conduct using this data and the types of affordances they can deploy as a result of their findings.

\textbf{Outcomes:} The consent assembly gave conditional consent. It was determined that assembly members generally thought sharing data, especially if it was deemed to improve accessibility features and enhance the functionality of websites, was appropriate. They thus agreed that they would recommend consenting by default to UX improvement purposes across all websites. However, assembly members were skeptical about using UX improvement purposes for running online research experiments on users, especially if it involved delivering ads. They were also concerned about the potential for UX improvement data to be misused in the design of addictive interfaces. They decided to opt out of consenting to these aspects of the purpose.

\textbf{Keeping Consent Decisions Updated:} With updates to data privacy regulations, industry data collection techniques, new accessibility features for websites, and new website functionalities, another consent assembly will need to reconvene to decide whether and how to update their consent decision for UX improvement purposes. Companies may in the future try to get consent from a new assembly if they require consent not covered by the old assembly’s output (e.g., because of fundamentally new functionalities, changing opinions, or technical capabilities).

\subsection{Vignette 2: Future Reuse of Consent and Data for GenAI Model Training} 
A new GenAI start-up has asked several online platforms for access to images their users have uploaded to their respective platforms. To train the GenAI model, the start-up will offer these platforms payment. Users uploaded these images on the platforms for a variety of reasons, ranging from sharing on social media to building websites. The companies will only share uploaded images from users of these platforms who originally consented to purposes such as ``sharing data with third parties'' in the context of the original platform they signed up for.

However, when originally consenting, these users could not predict that in the future, a new technology that generates new content from things they posted online would exist, and therefore did not consider this when initially providing their consent. Additionally, as consent notices are overwhelming, many users did not meaningfully familiarize themselves with the purposes when they made an account.

For the platforms, it is difficult to ask for individual consent to share these images with the start-up because tracking down \textit{who} to ask is nearly impossible. For instance, it could be possible that users who deleted their accounts had the model training on their images and videos. Many users commonly post photos of other people who were not tagged in the photo and thus were unable to consent to this photo sharing initially, and it is unknown exactly which data points from their users and non-users were collected. 

The platforms that want to sell user-uploaded images to the start-up have faced intense user and media scrutiny. They are unsure whether users' previous consent applies to feeding GenAI systems and want to avoid potential legal risk; therefore, they have requested a consent assembly to see if they can come to an agreement with the assembly. 

There is a lack of existing infrastructure to navigate collective consent for privacy in AI scenarios, but the national data protection authority has been tasked with facilitating this consent assembly as it touches on topics related to privacy and personal data.

Previously, the data protection authority sent a survey to several thousand people asking them about whether they used any of these platforms in the past and uploaded photos. A random selection of those who selected ``yes'' (to using a platform and uploading photos) and those who said ``no'' were invited to participate in the consent assembly.

\textbf{Inviting and Selecting Consent Assembly Members:} The consent assembly needs to consider the connected nature of online data, including images. For instance, users may post videos or images of others who do not have an account, and this content could all be potentially used to train a GenAI model. Therefore, this consent assembly should involve users of the original platforms who have uploaded photos in the past, as well as those who have not used these platforms or uploaded photos to them in the past. This mix of members within the consent assembly gauges users' general concerns and opinions are used for training GenAI on images uploaded to websites. 

It is important to also include others who have been historically impacted by AI systems~\cite{buolamwini2018gender} as part of the consent assembly process to mitigate the potential harms. In addition to the general assembly, the data protection authority ran a stratified consent assembly, this time involving women and racialized people, given their history of marginalization and the impact of AI systems on their communities.

\textbf{Learning and Listening:} AI systems, such as GenAI, are complex, rapidly evolving, and not well-understood by most people~\cite{storey2025generative}. Therefore, the data protection authority determined that for consent assemblies related to the reuse of uploaded images to train GenAI models, an in-depth deliberative mini-public (such as a citizens' assembly) could be organized to allow for this in-depth learning and exchange of ideas.

The assembly was presented with information from the company seeking to buy the different platforms' images, and the platforms that are considering selling their user images. They explained their data practices, training methodologies, and intended use cases. Additionally, the data protection authority determined that involving independent legal and technology experts in this consent was highly important. Independent legal experts provided perspectives on regulatory frameworks and users' rights, while technology experts offered an objective analysis of the technical capabilities, limitations, and risks associated with a positive decision. This multi-perspective approach ensures that assembly members receive balanced information rather than being exposed solely to industry viewpoints. 

Consent assembly members varied in their level of knowledge about GenAI. Similar to the wider population, some individuals were less informed and required additional learning sessions to understand this technology and how it worked. Assembly members had numerous questions for companies and other experts as they worked to understand the full scope of the issue. Some questions they asked include: How exactly are the uploaded images being collected and processed? How are GenAI models trained on this data, and what outputs can these models produce? What safeguards exist to protect privacy and prevent misuse? What are the potential benefits and harms associated with these practices? The assembly format provides dedicated time and space for these questions to be asked, discussed, and answered thoroughly, enabling participants to make informed consent decisions.

\textbf{Deliberation:} Based on the information presented in the learning and listening phase, the consent assembly can deliberate on whether they consent to the reuse of data and training GenAI models on social media images and videos. As decisions are collectively made by users and non-users, the assembly can make a representative decision not only for individual users who are directly implicated (i.e., have an account and uploaded images), but for others who may have had their data shared but could not consent.

\textbf{Outcomes:} This consent assembly rejected consent. Many participants disagreed with reusing previously uploaded images to train GenAI systems, stating that the original consent was not valid for this purpose, as users were unaware that this new technology could arise in the future when they initially consented, and had many ethical concerns about reusing data that was previously collected. Thus, the consent assembly did not consent to the reuse of images to train GenAI systems.

\textbf{Keeping Consent Decisions Updated:} In this particular scenario, if companies were to start anew and request another consent assembly for images uploaded to their platforms from a time when users are aware of GenAI's existence, another consent assembly will re-convene. This time, discussions regarding consent may focus on the types of websites to which consent is given for reusing uploaded images to train GenAI systems, or on which image types can and cannot be reused for GenAI purposes.

\section{Future Backcasting: Areas of Necessary Transformation}
\label{sec:backcasting}
We apply future backcasting, a methodology for futuring~\cite{servicedesigntools_futurebackcasting, vergragt2011backcasting, bendor2021looking}, to understand what is needed to change in our present day to allow for collective consent. Given the potentially controversial shift from individual to collective consent, we identified four key stakeholder groups (regulators and policymakers, businesses, civil society, and users) and developed strategic arguments to convince these stakeholders and make our speculative future (collective consent) a potential reality. 

\subsection{Legal Feasibility of Collective Consent}
\textbf{Stakeholder:} Regulators and Policymakers

\noindent \textbf{Why they need convincing:} When proposing fundamental changes to how consent is collected, it is reasonable to wonder how feasible these changes are, given that privacy regimes such as the GDPR and CCPA are well-established, have specific requirements for consent, and have set guidelines for what consent notices should look like to collect meaningful consent. Switching to collective consent would require overhauling these current institutions in place and replacing or significantly reworking them, which will result in investments of time and money.

\noindent \textbf{Why collective consent could work:} There was nothing specifically mentioned in the GDPR, CCPA, or other privacy regulations that consent and opting out had to be done using consent notices and similar banners. Instead, organizations started using these banners to comply with consent requirements, and the regulation adapted to these consent notices. 

Since consent notices were never officially recommended as the standard for consent, yet society and regulators adapted to this change, we believe there is space for reimagining consent. Lawmakers support revising regulations and their implementation to promote growth and innovation. In the UK, the Information Commissioner's Office has pledged to ``Relaxing enforcement of consent rules for privacy-preserving online advertising, ahead of exemptions to these legal requirements being introduced by government, where appropriate''~\cite{hm_treasury}. More recently, in November 2025 the European Union put forth a controversial Digital Omnibus in an attempt to reduce regulatory complexities, including changes to how consent could be implemented~\cite{Hart_Europe_2025}. These changes show that there are opportunities for new models of consent to be considered, and that it is time to make fundamental changes to online consent. 

The CCPA also allows authorized agents to fulfill data requests on a user's behalf~\cite{CCPA}, demonstrating that collective consent can feasibly work within some existing regulations. Collective consent decisions, informed by consent assemblies, could be recorded as GDPR Article 40 codes of conduct, which would require an Article 41 enforcement mechanism. This could provide a progressive path to their deployment. These decisions could serve as recommendations for enforcement authorities to establish basic defaults for consent, taking into account factors such as the purpose, website, and other relevant considerations to be determined. 

\noindent \textbf{Target:} The goal for regulators and policymakers would be that in the future, entities could be established, whether a branch of the public sector, the private sector, or a combination of both, to organize collective consent assemblies. These consent assemblies could be combined with internet standards bodies to establish internet standards that create default consent choices in the future. 

Under the GDPR, which relies on legal bases for processing user data, it may be possible to one day add a new legal basis for processing user data based on consent assembly decisions to further legitimize collective consent.

Consent is also largely viewed within the realm of data privacy, but many works have shown that consent intertwines with AI, healthcare, employment, and other domains; therefore, regulators in charge of these areas must be involved in collective consent, not just data privacy regulators. Additionally, consent compliance and enforcement have largely focused on consent notices and the development of design guidelines; however, regulators and policymakers need to adopt a broader view of consent beyond notices and interfaces, for collective consent to function properly.

One area that especially needs more development is infrastructure for conducting consent assemblies in the context of AI. AI governance lags behind the development of AI capabilities in many jurisdictions. The most established AI regulation is the EU AI Act, which is overseen and enforced by the European AI Office. However, the EU AI Act does not regulate the training of AI systems to the same degree as it regulates the deployment of AI systems~\cite {kyi2025governance}. Therefore, the establishment of new interpretations is needed to handle the implications of consent on GenAI data collection.

\subsection{Balancing Business Interests with User Privacy}
\textbf{Stakeholder:} Businesses

\noindent \textbf{Why they need convincing:} The internet of today is a transaction between companies and users. In exchange for using an online service, typically at no cost, the businesses collect user data to generate revenue. Collective consent may face challenges from companies because their outcomes could make it more difficult for companies to collect data. However, we also believe collective consent presents several benefits for companies.

\noindent \textbf{Why collective consent could work:} The first benefit of collective consent is that it offers simplified compliance. The EU and UK are proposing to relax consent requirements to reduce regulatory red tape~\cite{hm_treasury, Hart_Europe_2025}, showing that navigating complex regulations on the industry end is complex and a barrier to innovation. Additionally, several papers have shown that compliance with privacy regulations is a difficult task for organizations~\cite{horstmann2024those, kyi2025turning}, which has made tools like consent management platforms (CMPs) appealing for most companies to adopt~\cite{nouwens2020dark}. However, CMPs have also been found to be non-compliant with the GDPR~\cite{toth2022dark, nouwens2020dark}, adding more difficulty to the task of regulatory compliance. 

As collective consent is a process that is envisioned to have regulator involvement and oversight, there is less need for companies to worry about non-compliance, hiring lawyers to interpret the regulations, or dealing with the consequences of non-compliance because of misinterpretations. Being able to directly apply the recommendations for data collection set by a consent assembly could be an incentive for companies, as it simplifies regulatory compliance.

The second benefit is that collective consent can allow for more (regulated) data collection. With the introduction and enforcement of consent notices that allow users to reject all tracking, users are opting out of cookies. While this is beneficial from a privacy perspective, we must remember that businesses have an incentive to collect user data to operate their services and generate revenue. As a result, ad-tech companies are now dealing with lower consent rates, particularly for purposes essential to delivering advertising. Therefore, a default set by a consent assembly may not be as profitable for businesses as consent notices where users accept all cookies, but it may be more profitable than collecting no data from many users. 

\noindent \textbf{Target:} We envision collective consent as a balanced approach that protects the interests and rights of users while also offering a simpler approach to compliance, one that understands the business interests that sustain online services (typically at no cost).
To accomplish this, just like how businesses started putting consent notices on their websites, businesses could be required to apply collective consent decisions on their own websites to control how data is collected and used in the future. 

\subsection{Rethinking Personal Data Governance as an Individual to a Collective Problem}
\textbf{Stakeholder:} Society

\noindent \textbf{Why they need convincing:} While collective governance and bargaining are widely accepted in the workplace (such as through trade unions), the collective governance of personal data is still not widespread. This is because at the workplace, it is assumed that workers, due to their shared employer and employment conditions, have similar concerns, and thus a similar negotiating counterparty. In the context of privacy regulations, in particular the GDPR, they tend to be based on a human rights framework, under the belief that privacy (and thus data protection) is a fundamental human right~\cite{edps_dataprotection}.
As such, privacy becomes an individual right because human rights are inalienable rights unique to each person~\cite{un_humanrights}.

\noindent \textbf{Why collective consent could work:} We challenge the notion that personal data governance is an individual problem, and believe it would benefit users more if it were treated similarly to collective action in the workplace. As the internet has become monopolized by a few large technology companies~\cite{farrell2024we}, it is evident that users are subjected to shared online experiences and online harms~\cite{wu2025design}. Users of the internet face similar privacy concerns~\cite{wu2025design, clarke1999internet}, and similar concerns about what big platforms are doing with our data~\cite{kyi2024doesn, wu2023slow}. 

In a World Wide Web Consortium (W3C) report by \citet{BerjonYasskin2025}, it is also noted that "collective issues in data require collective solutions," stating that data governance cannot properly govern data if it focuses too much on individual control rather than collective action. 

\noindent \textbf{Target:} We believe that to make effective regulations that safeguard users' privacy, we need to shift from thinking of personal data governance as an individual problem, towards thinking of it as a collective one. This will require significant changes in the privacy regulations of many jurisdictions, as well as a societal shift in how we view personal data. The implementation of collective consent and governing personal data as a collective problem will also require new or significantly revamped institutions, as the current ones are often inadequate to handle the modern-day impacts of online platforms.

\subsection{Convincing Users About the Benefits of Collective Consent}
\textbf{Stakeholder:} Users

\noindent \textbf{Why they need convincing:} Adopting collective consent requires shifting users' perspective of data privacy and consent from individualistic to one that embraces shared decision-making. While the notion of individual consent is deeply ingrained in our society, existing frameworks and real-world examples can be leveraged to make the concept of collective consent more natural. When informing users about collective consent, it is essential to emphasize how this approach can lead to more efficient, equitable, and secure outcomes.

\noindent \textbf{Why collective consent could work:} The main obstacle to collective consent being accepted on the users' side is the idea that consent is an individual act. To overcome these beliefs, we can draw parallels to situations where individuals already delegate their decision-making power for the greater good. Democracy is a familiar example. Citizens elect representatives to make decisions on their behalf, trusting that these individuals will act in the best interests of the community. Similarly, in a union, workers collectively bargain and agree to abide by the decisions made by their union representatives.

Collective consent can be viewed as a form of "digital democracy" or "digital unionization," borrowing ideas such as delegating decision-making to others who are better equipped to make informed and meaningful decisions. Instead of every individual having to make a decision every time, a trusted group can manage it, freeing up others' time and cognitive load.

Users must, however, feel confident that their interests are being adequately represented. The representativeness of consent assemblies is crucial to building trust and ensuring the legitimacy of the collective consent process. To achieve this, those organizing consent assemblies must clearly define what constitutes a representative group and establish transparent processes for selecting assembly members and ensuring accountability. This could involve using a variety of metrics, such as ensuring demographic diversity, including individuals with different levels of privacy concern, as well as random and fair selection. The process must also provide a clear mechanism and infrastructure for individual users to opt out of collective consent if they feel their interests are no longer being served by consent assemblies. By emphasizing that collective consent is not about giving up control entirely but rather about entrusting it to a representative group, we can build a foundation of trust.

\noindent \textbf{Target:} We envision a future where entities running collective consent assemblies have established rapport with users. This means having proper representation in consent assemblies, making decisions that users generally agree with, and ensuring that users feel that delegating their decisions to a collective is more effective than individual consent. 

%% file: content/06_discussion.tex
\section{Discussion}
\label{sec:discussion}
Using speculative design and future backcasting, we conceptualize how collective consent can be implemented to govern personal data more effectively. Collective consent proposes much-needed fundamental changes to consent not addressed by current approaches by recognizing that data is often connected, which has communal privacy implications. We envision that collective consent can also have the potential to apply to a variety of data collection scenarios beyond replacing notice and consent, and is centered on the ability for users to be truly informed, negotiate, have the option to opt out, and keep their consent decisions up-to-date. 
In this section, we discuss the broader implications of collective consent. 

\subsection{Collective Consent and Effective Consent}
Based on the criteria for effective consent presented in Section \ref{sec:effective_consent}, we discuss how collective consent fulfills each of these elements of effective consent, and areas for improvement and additional consideration:

\begin{itemize}
    \item \textbf{Freely given (GDPR)}: Consent assemblies are less susceptible to influence from businesses, as they are making decisions in groups facilitated by a neutral party, such as a regulator, who would have users' privacy as a primary interest. Collective consent also considers the societal context of consent, such as who can meaningfully consent, and acting collectively gives users more power. 
    \item \textbf{Specific (GDPR)}: The GDPR emphasizes individual choice and granular control, whereas consent assemblies involve collective decision-making that may not reflect each participant's independent preference and may struggle to provide the specific granularity GDPR requires, especially for different organizations. However, consent assemblies might function more compatibly with GDPR if reconceptualized as complementary mechanisms that inform organizational practices and help establish privacy-protective defaults or recommended consent options that individuals can then accept or modify, preserving individual decision-making authority while ensuring options reflect collective values and concerns.
    \item \textbf{Informed (GDPR)}: Deliberative mini-publics put emphasis on thorough learning and deliberation to understand a concept. Consent assemblies also adopt a similar model, allowing members to learn and deliberate on the topic to become better informed. 
    \item \textbf{Unambiguous (GDPR)}: Collective consent decisions will apply to the purposes for which they are being made. These decisions will be based on the purpose of data collection, but assemblies can establish conditions for consent based on the type of organization collecting data and the type of data being processed.
    \item \textbf{True user agency (World Economic Forum)}: Collective consent is envisioned as a process where the consent assembly can negotiate with organizations collecting data and allows for the ability to opt out. 
    \item \textbf{Participatory (World Economic Forum)}: Collective consent leverages deliberative mini-publics, which are a participatory form of democratic decision-making. Not only are consent assemblies participatory for users, but they are also envisioned to involve a jury of various stakeholders involved in consent, such as legal and technology experts. 
    \item \textbf{Real choice (World Economic Forum)}: Collective consent offers the ability for users to be truly informed of what they would be consenting to. Through the process of a consent assembly, users can make granular consent decisions, opt out without penalty, and negotiate the terms for consent. 
\end{itemize}

In addition to these elements, we also mentioned three ways notice and consent is failing in Section \ref{sec:background:ineffective}. We discuss how collective consent addresses these challenges, along with areas for further consideration and improvement:

\begin{itemize}
    \item \textbf{UI challenges}: Collective consent is not based on a user interface that users need to interact with. Instead, it relies on a representative group to make decisions, and this consent decision is then sent to organizations where it would be applied across a jurisdiction. 
    \item \textbf{Assumes every user can equally consent and opt out}: By having a representative consent assembly, including demographic-specific assemblies, there is better representation of users who can make meaningful and informed decisions for the collective. Consent assemblies are designed to enable users to opt out of data processing if they so wish. We recommend options for individual users to opt out of collective consent decision-making if they wish, and have the choice about whether to be represented by a collective or not. 
    \item \textbf{Consent for a wider range of data collection practices}: Collecting individual consent when the data is difficult to trace, when it has already been collected, or when there is a possibility that the data may be reused is very challenging. A collective approach to data governance may be more appropriate than individual approaches, as collective consent is centered on establishing default guidelines for consent, and accounts for the communal implications of data ecosystems. 
\end{itemize}

\subsection{Downsides of Collective Consent}
We recognize that collective consent, while envisioned to improve current challenges presented by notice and consent and its alternatives, is not a perfect solution and presents some downsides.

First, the focus on collective-level privacy comes with a decrease in individual privacy and autonomy compared to consent notices and other alternatives. There may be the erosion of individual choice by prioritizing the collective's choice, where the group's privacy becomes more important than individual privacy. To address this, some possibilities to explore include the option to allow users to opt out of collective consent decisions and to run consent assemblies for specific demographic groups, such as children and those with special needs, who may have different privacy considerations than the majority. 

Second, implementing collective consent may be time-consuming. Regulation, society, users, and industry have adapted to consent notices; however, we are calling for an overhaul of this model. To make collective consent more feasible to implement, it is worth exploring the possibility of conducting remote consent assemblies, leveraging AI (such as using an AI moderator). Previous papers by \citet{fishkin2019deliberative} and \citet{argyle2023leveraging} have shown promising results for online deliberative mini-publics that use an AI moderator. 

Third, just like consent notices have been subjected to manipulation via deceptive designs~\cite{nouwens2020dark, kyi2023investigating, machuletz2019multiple}, collective consent has the potential for manipulation. Companies may have a business interest in collecting more data, and may find ways to influence consent assemblies. To mitigate outside influence, consent assemblies must be organized and operated by a neutral party with no stake in any organization's interests, such as regulators or privacy-oriented NGOs. 

\subsection{Collective Consent for Other Technology Governance Scenarios}
We believe that collective consent can be effectively applied in other scenarios where privacy is a concern. Here, we list out some potential applications for collective consent, but note this is not an exhaustive list of possibilities:

\begin{itemize}
    \item \textbf{Content moderation:} Currently, companies have control over content moderation policies on their platforms, a concept that has been criticized and subject to change depending on business or political interests~\cite{Milmo2025, erickson2025content}. Although content moderation is not typically seen as something one consents to when using a platform, it impacts what a user is exposed to and may involve collecting user data to show personalized, moderated content. To make content moderation more reflective of the wider population, we envisage collective consent assemblies being applied to content moderation to understand what kind of content assemblies believe is appropriate to moderate or not on certain platforms, and how content moderation may be adapted over time in response to changes to society. 
    \item \textbf{Smart home:} Devices such as smart speakers and doorbell cameras are shared or managed communally, impacting the privacy of everyone living within a household, and those who visit a household (e.g., delivery drivers, friends, babysitter, etc.). This leads to several privacy concerns around smart homes~\cite{zeng2017end}. Individual consent cannot be collected, as many individuals are impacted by smart devices, and obtaining this consent may be challenging (e.g., due to the time constraints of delivery drivers who are in a rush to complete their deliveries). In this case, collective consent can be applied to understand the general sharing and privacy preferences surrounding different smart home devices. Based on the findings from the collective consent assembly, default privacy settings can be adapted to reflect who is interacting with the household (i.e., a resident or visitor) and the type of data the device collects. 
    \item \textbf{DNA data:} Genetic data not only reveals sensitive personal information about the individual giving their DNA sample, but also those related to them and the broader community they belong to, causing many privacy concerns~\cite{baig2020m}. Because consent for instances where data is linked cannot be collected on an individual basis, collective consent is necessary~\cite {walquist2025collective}. Collective consent could potentially come in the form of \textit{family-based consent}, where the family of the individual giving their DNA makes a shared decision together, or in the form of \textit{community-level governance} when larger-scale genetic projects are happening (such as collecting DNA from specific and underrepresented ethnic groups, or from a specific region).
    \item \textbf{Bystander Privacy:} Protecting the privacy of bystanders, individuals that are accidentally captured in public data, such as photos, videos, satellite imagery, or through wearable technologies, is important~\cite{zhangmore}. Securing explicit, individual consent is often impossible in these scenarios, as noted by \citet{zhangmore}, who call for ``establishing strong default privacy protections.'' Therefore, collective consent offers a viable alternative. This approach can establish agreed-upon social contracts or technical frameworks that govern the standards for capturing, storing, and using data about bystanders.
\end{itemize}

\subsection{Collective Consent: Towards Internet Standards}
A promising solution emerges through combining collective consent assemblies with Internet standards development. Under this framework, consent assemblies would play a crucial role in informing what internet standards should be, ensuring that these standards reflect actual user preferences rather than being dictated by regulators or industry alone. These consent assemblies must be grounded in user-recommended consent decisions, as regulators cannot and should not make these determinations independently without meaningful input from users.

This approach can offer the benefit of reducing the regulatory burden on authorities by decreasing the need to establish numerous consent assemblies in the future. By embedding privacy considerations directly into internet standards, the framework could streamline privacy governance while maintaining democratic legitimacy through collective input.

To ensure the effectiveness and legitimacy of this system, Internet standards organizations and their governing bodies must maintain ongoing dialogue with governments, industry, and users. This continuous engagement would help guarantee that the standards remain representative of users' interests and adapt to evolving privacy needs, technological and regulatory developments over time.

\subsection{Future Work}
As this is a theoretical paper on collective consent, further empirical work is needed in this area to understand how to effectively implement collective consent and empower users with collective privacy decision-making. Some future work could explore what users need to feel represented by a consent assembly, such as identifying the demographics that are important for representation, the characteristics of designing a consent assembly, and how to create a representative decision-making process. 

Work can also look into understanding the parameters for collective consent, such as how broad or narrow data collection purposes should be deliberated on. We also encourage further research to explore how collective consent can be applied to other forms of data governance and scenarios where privacy is a concern. Empirical work can provide a deeper investigation into collective consent in the context of examples we mentioned, such as smart home devices, DNA data, content moderation, or other use cases. Future work can also explore the legal dimensions of collective consent, such as how to implement it within pre-existing legal frameworks, such as the GDPR and CCPA, or how these legal frameworks could be adapted in the future to recognize group rights and community-level data interests.

\section{Conclusion}
Current approaches to consent are focused on the consent of an individual. This individualistic approach is not realistic for modern-day data processing, because user data is often interlinked, has communal privacy implications, and having every user properly informed of all data collection practices online is impractical. Therefore, a collective approach to consent could be a reasonable alternative for collecting users' meaningful and informed consent. 

Using a speculative design process, we conceptualize how collective consent, imagined through consent assemblies, could be implemented by leveraging deliberative mini-publics. We present two vignettes, i) replacing notice and consent notices, and ii) reuse of data to train GenAI models, to further illustrate how consent assemblies could be implemented. We apply future backcasting to understand what we need to change in the present to make collective consent a reality for regulators, businesses, society, and users in the future. We envision a future where consent is fundamentally transformed, empowering users to make more informed and meaningful decisions in a variety of technology governance scenarios, and collective consent could be the way forward.